%% file: ms.tex
\documentclass[12pt,preprint]{aastex}

\shorttitle{\pks{} Jet Low-Energy Electron Distr. Cutoff}
\shortauthors{Mueller \& Schwartz}

\newcommand{\gammin}{\gamma_{\rm min}}
\newcommand{\gammax}{\gamma_{\rm max}}
\newcommand{\gamcrit}{\gamma_{\rm crit}}
\newcommand{\nh}{n_{\rm H}}
\newcommand{\pks}{\mbox{PKS 0637--752}}
\newcommand{\axaf}{\emph{Chandra}}
\newcommand{\xray}{\mbox{X-ray}}
\newcommand{\etal}{\mbox{\emph{et al.}}}
\newcommand{\highnh}{$\nh = 9.1 \times 10^{20}$ cm$^{-2}$}
\newcommand{\scgammin}{\gammin \, \delta_{10}}

\begin{document}

\title{Constraints on the Low-Energy Cutoff in the Electron Distribution
of the \pks{} Jet}

\author{M. Mueller}
\affil{Stanford Linear Accelerator Center, 2575 Sand Hill Road, Menlo Park, CA 94025, \\
and Kavli Institute for Particle Astrophysics and Cosmology, Stanford University, Stanford, CA 94305}
\email{mmueller@slac.stanford.edu}

\and

\author{D. A. Schwartz}
\affil{Harvard-Smithsonian Center for Astrophysics, 60 Garden Street, Cambridge, MA 02138}
\email{das@head.cfa.harvard.edu}

\begin{abstract}
We re-analyze the \axaf{} ACIS spectrum of the kpc-scale jet in \pks{} to 
investigate the 
possible low energy cutoff in the relativistic electron spectrum 
producing the non-thermal radiation in the scenario of inverse Compton emission 
off the cosmic microwave background. 
This was among the first objects targeted by the \axaf{}
Observatory and gives a unique opportunity to study the low energy \xray{}
emission free of detector contamination. 
As previously reported in the literature, the spectrum can be fit by
a power law, with the slope predicted by the radio spectrum, modified by 
low energy absorption through the Galaxy as determined from the spectrum of the
quasar core and by HI 21 cm observations. 
We obtain a marginally better fit with an model of inverse 
Compton emission produced by an electron population that exhibits a cutoff at 
$\scgammin$ between about 50 and 80 (assuming $\Gamma = \delta$). This range for 
$\gammin$ is higher than has previously been assumed in broad-band spectral fits to 
the jet emission. 
The observed optical flux can be used to place a lower limit on $\gammin$; 
the constraint is not very strong, but does suggest that $\gammin$ must be 
higher than 1 to avoid overproducing the optical emission. 
We investigate the effect of uncertainties in the column density for galactic 
absorption as well as the calibration of \axaf{} for these early observations.
Finally, we discuss the implication of these limits on the jet luminosity 
in this source.
\end{abstract}


\section{INTRODUCTION}

The initial observations to focus \axaf{} using the presumed point
source \pks, a quasar at redshift 0.651 
\citep{savage76}, 
revealed a
remarkable \xray{} jet structure 
\citep{schwartz00,chartas00},
coincident with the radio jet reported by
\citet{tingay98}. 
\citet{schwartz00} 
pointed out that the \xray{}
emission could not be explained reasonably by thermal bremsstrahlung,
by an extension of the radio synchrotron spectrum, by synchrotron
self-Compton, or by inverse-Compton (IC) from any likely source of seed
photons provided that the magnetic field was near its equipartition
value. Nonetheless they concluded that IC from the same
electrons producing the radio emission was the most likely source of
the \xray s in view of the similarity of the \xray{} and radio surface
brightness profile.

\citet{tavecchio00} 
and 
\citet{celotti01} 
solved the dilemma by invoking bulk
relativistic motion of the \mbox{100 kpc} scale jet. Relativistic motion with
bulk Lorentz factor $\Gamma$ has the effect of requiring a smaller
magnetic field in the rest frame for minimum energy conditions, and of
increasing the apparent energy density of the cosmic microwave
background (CMB) by a factor of $\Gamma^2$ on average as seen by the radiating
electrons. Such an IC/CMB explanation subsequently has been
applied successfully to explain the \xray{} emission from
powerful quasar and \mbox{FR II} radio jets; e.g., by 
\citet{sambruna02,harris02,siemiginowska02,siemiginowska03,sambruna04,marshall05,schwartz06,sambruna06} 
and
\citet{schwartz07}.

\citet{tavecchio00} 
pointed out that the scenario in which low energy 
electrons underwent Compton scattering on an external photon field 
gave an opportunity to infer parameters of the low energy portion of the 
relativistic electron distribution for the first time. They estimated 
$\gammin \sim 10$ for the \pks{} jet, by constructing a fiducial 
spectral model which 
fit the broad-band radio, optical and \xray{} flux data. Subsequent examinations 
of the jet spectrum 
\citep{celotti01,uchiyama05} 
all converged on a value for $\gammin \sim 10$--20. In the present 
work we perform a more rigorous assessment of $\gammin$  by performing 
a detailed spectral fitting of the ACIS \xray{} spectrum. 
\pks{} is in the unique position among identified \xray{} jet sources 
in that, because the exposures were done so early in the lifetime of 
\axaf , the low-energy \xray{} spectrum is free of the detector contamination and 
associated loss of sensitivity that affect later observations. 
The main source of uncertainty remains the low-energy calibration of 
ACIS, and we contrast results obtained from a wider bandpass 
(0.3--7 keV) than 
is usually used for imaging spectral analysis using \axaf{} with 
the equivalent results if the bandpass is restricted to the better-calibrated 
range of 0.55--7 keV.

\mbox{Section 
\ref{sec:observations}} 
lists the observations and data reduction for this investigation. 
The development of the custom IC/CMB spectral models is detailed 
in \mbox{Section 
\ref{sec:iccmb_model},} 
which are then applied to the data in \mbox{Section 
\ref{sec:results}} 
to derive limits on $\gammin$. The interpretation of the model fits, including 
a comparison between two different prescriptions for the calculation of 
the IC/CMB spectrum and a discussion of the effect of the shape of the electron 
distribution at the $\gammin$ cut-off, is the topic of \mbox{Section 
\ref{sec:discussion}.}
Section 
\ref{sec:conclusion} summarizes the paper.

\section{OBSERVATIONS AND DATA REDUCTION}
\label{sec:observations}

\pks{} was observed by \axaf{} using the \textsc{acis-s} 
instrument as part of the initial focusing period. This early in the 
mission, there was no loss of low-energy sensitivity due to the 
contamination layer. We thus have an opportunity with this object to 
push the \xray{} spectrum to lower energies than is currently possible.

\mbox{Table 
\ref{tbl:observations}} 
lists the details of the \axaf{} observations that 
were used in this investigation, all of which were carried out in the 
\textsc{faint timed} mode of \textsc{acis}.  We selected the 
observations based on whether the quasar and jet were positioned on 
the back-illuminated S3 chip, which excludes two observations (ObsID 
1059 and 1061).

Other observations were discarded: ObsID 1057 and 62558 
because the identification of the quasar and jet was unclear, and 
ObsID 1093, 1264, and 1265 since they do not appear to contain any useful data. 
Although most of the remaining 
observations were not at the best focus, the quasar and jet are 
clearly separated, and the jet spectrum can be extracted without contributions 
from the quasar core.

For each observation, the evt2 file was re-extracted from the evt1 file in order to 
apply the latest calibrations, as outlined in the ACIS data preparation thread on 
the CIAO web page (CALDB version 3.2.4). 
Since the observations were taken early in the mission, and the focal temperature was 
different from $-120\,^{\circ}$C, no time-dependent gain or CTI correction was used. 

\begin{figure}
\plotone{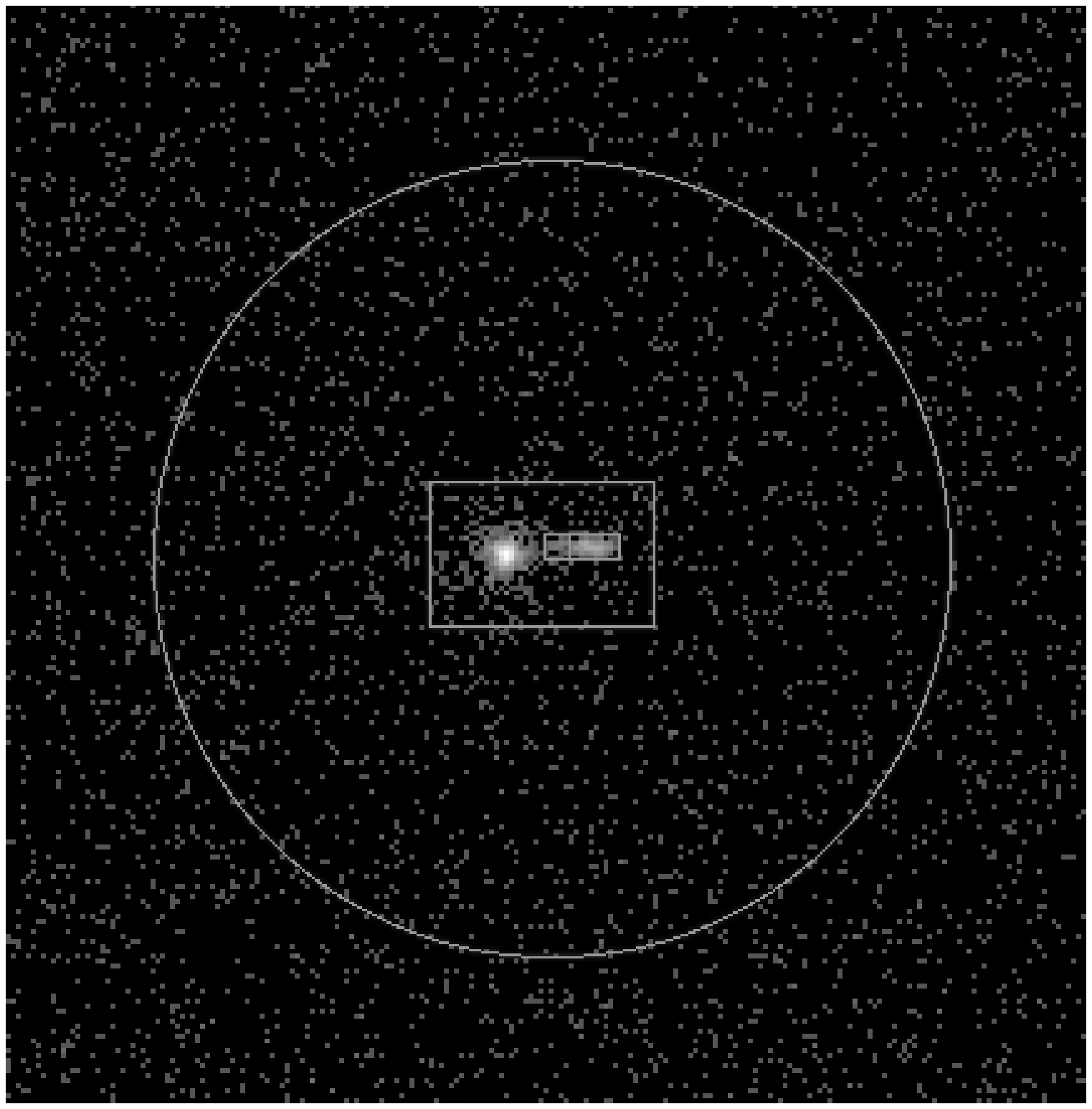}
\caption{Extraction regions superimposed on one of the exposures 
of \pks{} (ObsID 62554). The inner and outer jet regions are the 
two rectangular regions straddling the jet; the background was 
extracted from the elliptical region while excluding the counts from 
within the larger rectangular box. These extraction regions are 
representative of the regions used for all observations; however, 
the change in focus for some of them necessitated larger boxes 
for the inner and outer jet, which were then chosen to again just 
straddle the image of the jet.
\label{fig:extraction_regions}}
\end{figure}

Extraction regions for the jet and background 
were defined individually for each observation, since in some cases the departure from 
optimal focus necessitated a larger extraction region to include all events from the 
jet. The extraction regions were divided into an inner and outer jet, 
the first of which included the faint \xray{} emission of the bridge between the 
quasar core and the bright radio/optical/\xray{} knots, which comprised the second 
region. The regions for one of the observations (ObsID 62554) are shown in Figure 
\ref{fig:extraction_regions}. 
No significant differences in the spectra between these two sub-regions 
were found, although the low number of counts in the bridge spectrum make a detailed 
comparison difficult. All subsequent fits used only the outer jet extraction region.

Using the CIAO tool \texttt{dmextract}, the counts in each observation's source and background 
extraction regions were combined into pulse invariant (PI) channels to form the individual spectra. 
A grouping algorithm was applied to have a minimum of 2 counts per bin. Weighted ancillary 
response files (WARF) and redistribution matrices (WRMF) were created using the tools 
\texttt{mkwarf} and \texttt{mkrmf}, as appropriate for extended sources. (\texttt{mkacisrmf} 
does not apply for these observations due to their early observation dates.)
To check for 
background flares, the light curve of the S3 chip excluding the region around 
the quasar and the jet was analyzed visually. Data during times corresponding to 
the few occurrences of spikes in the count rate were filtered out in the extraction.

Because the results reported in this investigation depend heavily on the correct modeling 
of the low-energy response of \texttt{acis}, alternative calibration products were extracted 
to investigate the effect of uncertainties in the calibration 
(Grant, C. \& Bautz, M., private communication). Sets of WARF 
and WRMF were constructed using $-100\,^{\circ}$C and $-120\,^{\circ}$C FEF files. In 
addition, for the $-100\,^{\circ}$C FEF calibration products, the Quantum Efficiency 
Uniformity file was varied between the flat version N0001 and the version N0002 that 
includes information on how the quantum efficiency varies over the S3 chip. 
Spectral fits were repeated with both the nominal set of calibration files and 
the above alternative calibrations. The results on $\gammin$ are unaffected to within the 
uncertainties introduced by the choice of IC scattering kernel and the shape of the 
$\gammin$ cutoff (see Section 
\ref{sec:iccmb_model}). 
However, we will report results obtained from both a 0.3--7 keV bandpass as well as a 
better-calibrated 0.55--7 keV bandpass.

\pks{} was also observed by the \emph{Hubble Space Telescope} 
\citep{schwartz00}. The area where the three optical knots are located corresponds 
morphologically to the extraction region used for the \xray{} analysis. The total 
optical flux density from the knots is $0.574 \mu$Jy at a frequency of 
$4.3 \times 10^{14}$ Hz.

\section{IC/CMB MODEL}
\label{sec:iccmb_model}

We wish to fit the spectrum for the jet with a model of inverse Compton scattering of energetic 
electrons on the CMB. 
We assume that the spectrum of the entire \xray{} jet is emitted by a spatially homogeneous 
population of 
electrons that collectively exhibit a bulk motion with Lorentz factor 
$\Gamma = (1 - \beta^2)^{-1/2}$. The 
random motion of the electrons (assumed isotropic in the jet rest frame) is described by 
the distribution of the Lorentz factor $\gamma$, such that the energy of any given electron 
in the jet rest frame is $\gamma m_e c^2$ ($m_e$: electron rest mass, $c$: speed of light; 
primed quantities refer to the jet rest frame, unprimed to the observer frame; exception: $\gamma$, 
which always refers to the jet rest frame).

This distribution is modeled as a power law between $\gammin$ and 
$\gammax$, with a slope $s$ ($n'(\gamma) \propto \gamma^{-s}$). 
The upper end of the distribution is chosen as a sharp cutoff; since we note 
that the high energy cutoff is not reached in the \xray{} spectrum, the exact functional form 
doesn't affect the analysis as long as $\gammax$ is chosen sufficiently 
high. The shape of the low-energy cutoff, however, has a direct effect on the 
model spectrum. We investigate two options: a step function at 
$\gammin$: 

\begin{equation}
n'(\gamma) = \left\{ \begin{array}{ll}
N_0' \, \gamma^{-s}   &   \gammin \leq \gamma < \gammax\\
0                    &   \textrm{otherwise,}
\end{array} \right.
\label{eq:e_distr_sharp}
\end{equation}

and a constant electron density below $\gammin$:

\begin{equation}
n'(\gamma) = \left\{ \begin{array}{ll}
N_0' \, \gammin^{-s}   &   \gamma < \gammin\\
N_0' \, \gamma^{-s}    &   \gammin \leq \gamma < \gammax\\
0                     &   \gammax \leq \gamma.
\end{array} \right.
\label{eq:e_distr_flat}
\end{equation}

Neither of them is likely to be the actual shape of the low-energy cutoff, but they 
(together with the case when $\gammin$ is sufficiently small as to move the cutoff 
outside of the spectrum bandpass) provide three representative cases for the 
characterization of the electron distribution based on the available \xray{} data.

$j'( E'_1, \Omega'_1 )$, the IC/CMB emissivity in the jet rest frame at a given energy $E_1'$ and 
direction $\Omega'_1$, can be formulated as an integral over the invident electron and 
photon energies and their directions, taking into account their respective number densities (which 
are themselves functions of the corresponding particle's energy) and 
the relativistic transformations of the energies and directions between the jet rest frame and the 
individual electron's rest frames. The scattering of CMB photons into the \xray{} band is due to 
electrons with Lorentz factors small enough to make the scattering event in the electron rest frame 
be safely within the Thomson limit. 
This and other assumptions allow these integrals to be evaluated 
in terms of elementary functions. We wish to compare two different 
approaches that differ in the way the photon field is treated.

\subsection{Blumenthal \& Gould Approach}

The 
\citet{blumenthal70} 
calculation assumes an isotropic distribution of seed photon directions, which 
is strictly speaking not applicable to a jet moving relativistically with respect to the CMB 
rest frame. However, the simplicity of the resulting expression for the emissivity is the reason 
it has been used extensively in the literature. Since both the electron and photon fields are treated 
as isotropic, the emissivity does not depend on $\Omega'_1$ and can be written as 

\begin{equation}
j'( E'_1 ) \propto E'_1 \int_{\gamcrit}^{\gammax}
\int_{E'_0}
\frac{n'(\gamma)}{\gamma^2} \, f(E_0', E_1', \gamma) \, \mathrm{d}E'_0 \, \mathrm{d}\gamma ,
\label{eq:blugou_emissivity}
\end{equation}

where $f(E_0', E_1', \gamma) = 2 x \ln x - 2 x^2 + x + 1$, with $x = E_1' / (4 \gamma^2 E_0')$. 
$\gamma \gg 1$ 
is assumed throughout. The function $f$ is often called the kernel of IC scattering, because it 
describes the spectrum obtained from a fixed initial electron and photon energy. 
$\gamcrit$, which is the Lorentz factor of 
the least-energetic electrons that can contribute to the emission at $E'_1$, is derived as 

\begin{equation}
\gamcrit = \frac{1}{2} \, \sqrt{\frac{E_1'}{E_0'}}.
\end{equation}

A further simplification 
treats the incident CMB photon field as monoenergetic at an energy 
$E'_0 = (1 + z) \, \Gamma \, \sigma \, k T_{\rm CMB}$, 
where $\sigma = 2.7$ is the average photon energy of a thermal spectrum in units of $kT$
\citep{felten66},
$k$ is the Boltzmann constant, and $T_{\rm CMB} = 2.725$ K is the local CMB temperature 
\citep{fixsen96}. 
The integral over $E'_0$ in Equation 
\ref{eq:blugou_emissivity} 
is thus eliminated.
The motivation for this simplification is that features in the observed \xray{} spectrum 
are expected to be much broader than the width of the thermal distribution of CMB photon energies.

In the case of the electron distribution with the sharp cut off at $\gammin$ (Equation 
\ref{eq:e_distr_sharp}), 
the final expression for the emissivity is 

\begin{equation}
j'( E'_1 ) \propto E'_1 \int_{\mathrm{max}(\gamcrit,\gammin)}^{\gammax}
\gamma^{-s-2} \, f(x) \, \mathrm{d}\gamma .
\end{equation}

For the other electron distribution (Equation 
\ref{eq:e_distr_flat}), 
the equivalent expression is

\begin{equation}
j'( E'_1 ) \propto E'_1 \int_{\gamcrit}^{\gammax}
\gamma^{-s-2} \, f(x) \, \mathrm{d}\gamma ,
\end{equation}

if $\gamcrit \geq \gammin$, and 

\begin{equation}
j'( E'_1 ) \propto E'_1 \int_{\gammin}^{\gammax}
\gamma^{-s-2} \, f(x) \, \mathrm{d}\gamma + E'_1 \, \gammin^{-s} \, \int_{\gamcrit}^{\gammin}
\gamma^{-2} \, f(x) \, \mathrm{d}\gamma ,
\end{equation}

if $\gamcrit < \gammin$.

The observed spectral flux $j(E_1)$ is proportional to the emissivity in the 
jet rest frame $j'(E_1')$ when the energy shift of the photons due to the motion of the 
jet rest frame with respect to the observer is taken into account. 
Photons emitted at an energy $E_1'$ in the jet rest frame are observed 
at an energy 
$E_1 = \delta \, E_1' / (1 + z)$ in the observer frame, where 
$\delta = [\Gamma (1 - \beta \mu)]^{-1}$ 
is the jet Doppler factor (with $\theta = \cos^{-1} \mu$ the observer 
viewing angle of the jet) and $z$ is 
the redshift of the quasar. 
Since for the determination 
of $\gammin$ we are only 
interested in the shape of the spectrum, and not its normalization, the implementation 
of the model simply uses the jet rest frame emissivity appropriately shifted along the 
energy axis and applies a normalization factor that is not specified in detail to best fit 
the observed number of counts.

\subsection{Aharonian \& Atoyan Approach}

In contrast to 
\citet{blumenthal70}, 
the 
\citet{aharonian81}
approach treats the incident photon field as monodirectional (anti-parallel 
to the jet bulk velocity for IC/CMB) in the jet rest frame, which is an appropriate approximation 
for $\Gamma \gg 1$. Because the electron distribution in the 
jet rest frame is assumed isotropic, 
the system now exhibits azimuthal symmetry 
around the propagation direction of the jet, and the angular dependence of the emissivity 
reduces to a dependence on the polar angle $\theta' = \cos^{-1}\mu'$ of the outgoing photon.
If $\gamma \gg 1$ is assumed as well, 
then, irrespective of the scattering angle in the electron rest frame, the direction 
(in the jet rest frame) of the 
photon after scattering is well approximated by the direction of the 
incident electron. The integration over the incident electron direction thus reduces 
to a $\delta$-function substitution.

Adopting the notation of 
\citet{stawarz05}, the emissivity can be written as 

\begin{equation}
j'( E'_1, \mu' ) \propto E'_1 \int{\gamcrit}^{\gammax} \, \mathrm{d}\gamma \,
\int \, \mathrm{d}E'_0 \,
\frac{n'(\gamma)}{\gamma^2} \, f(E_0', E_1', \gamma, \mu') ,
\label{eq:ahaato_spectrum}
\end{equation}

with the IC kernel, expressed using the quantities 
$v' = 2 \, (1 - \mu') \, E_0' \gamma / (m_e c^2)$ and 
$w' = E_1' / (\gamma m_e c^2)$, 

\begin{equation}
f(E_0', E_1', \gamma, \mu') =   1 \, + \, \frac{w'^2}{2 \, (1 - w')} \, - \, \frac{2 w'}{v' \, 
                                    (1 - w')}
                                + \, \frac{2 w'^2}{v'^2 \, (1 - w')^2} .
\end{equation}

In this case, $\gamcrit$ evaluates to

\begin{equation}
\gamcrit = \frac{E_1'}{2 m_e c^2} \left\{ 1 + \left( 1 + \frac{2 m_e^2 c^4}{(1 - \mu') \, E_0' \, 
E_1'} \right)^{1/2} \right\}.
\end{equation}

Again, the condition that $\gamma \gg 1$ has to be satisfied. The additional condition in 
\citet{aharonian81}, 
that $E_1' \gg E_0'$, is the same simplifying assumption mentioned in the 
explanatory text to Equation 7.26a in 
\citet{rybicki} and to Equation 2.13 in 
\citet{blumenthal70} 
and is readily satisfied for IC/CMB \xray s. While the above expression is derived in the 
Klein-Nishina regime, the authors stress that the exact formula for the IC spectrum (without 
the simplifying assumptions mentioned above) is valid for any values of the electron and 
photon energies. Since the subsequent simplifications are satisfied in the Thomson regime also, 
Equation 
\ref{eq:ahaato_spectrum} 
is applicable to the process of \xray{} generation via IC/CMB. In addition, $f(E_0', E_1', \gamma, \mu')$ 
does not appear to require any modifications to mitigate potential numerical precision issues in its evaluation 
in the Thomson limit.

In the case of the electron distribution with the sharp cut off at $\gammin$ (Equation
\ref{eq:e_distr_sharp}), 
the final expression for the emissivity is 

\begin{equation}
j'( E'_1, \mu' ) \propto E'_1 \int_{\mathrm{max}(\gamcrit,\gammin)}^{\gammax}
\gamma^{-s-2} \, f(E_0', E_1', \gamma, \mu') \, \mathrm{d}\gamma .
\end{equation}

For the other electron distribution (Equation 
\ref{eq:e_distr_flat}), 
the equivalent expression is

\begin{equation}
j'( E'_1, \mu' ) \propto E'_1 \int_{\gamcrit}^{\gammax}
\gamma^{-s-2} \, f(E_0', E_1', \gamma, \mu') \, \mathrm{d}\gamma ,
\end{equation}

if $\gamcrit \geq \gammin$, and 

\begin{equation}
j'( E'_1, \mu' ) \propto E'_1 \int_{\gammin}^{\gammax}
\gamma^{-s-2} \, f(E_0', E_1', \gamma, \mu')  \, \mathrm{d}\gamma + E'_1 \, \gammin^{-s} \, \int_{\gamcrit}^{\gammin}
\gamma^{-2} \, f(E_0', E_1', \gamma, \mu') \, \mathrm{d}\gamma ,
\end{equation}

if $\gamcrit < \gammin$.

As in the 
\citet{blumenthal70} 
approach, the observed spectrum $j( E_1, \mu ) \propto j'( E'_1, \mu' )$.
$\mu'$ is related to $\mu$ and $\beta$ 
through $\mu' = (\beta - \mu) \, / \, (1 - \beta \mu)$, 
when the incident photons in the jet rest frame are traveling in the opposite direction 
to the jet. 
Since the jet viewing angle is small 
\citep{lovell00}, 
we assume that $\Gamma = \delta$, which implies $\mu = \beta$, and thus 
$\mu' = 0$.

\subsection{Model Implementation}

We proceed to implement custom \textsc{xspec} 
\citep{arnaud96} 
models that compute spectra 
based on the above formulas. The numerical integration is done using the 
\texttt{qsimp} routine in 
\citet{Press}. 
For both approaches, the two adjustable model parameters are $s$ (the electron 
distribution power law index)
and $\gammin$. The additional parameters that are not adjusted 
in the fit are $\gammax$, $\Gamma$, $\delta$, and $z$. 
$\gammax$ is kept fixed at $10^5$ (corresponding to a high-energy cut-off around 
850 MeV); $z = 0.651$. As mentioned before, the normalization of the model 
is arbitrary, as the correspondence between the normalization and the relevant 
physical quantities like luminosity and distance is not spelled out.

Following 
\citet{tavecchio00} and 
\citet{schwartz07}, 
we assume $\Gamma = \delta = 10$, which sets $E_0' = 1.05 \times 10^{-5}$ keV. 
From the $\delta$-function approximation to the energy spectrum of IC scattering 
from a fixed initial electron and photon energy (where $E_1' = 4/3 \, \gamma^2 \, E_0'$ fixed), 
the Lorentz factor for electrons scattering 
CMB seed photons to a given observed energy scales as 
$(\Gamma \, \delta)^{-1/2}$. Under the assumption that $\Gamma = \delta$, this reduces to 
a scaling by $\delta^{-1}$. The same behavior is expected for the two kernels used in the 
models, although this was not investigated in detail. 
Our quoted results for $\gammin$ are thus expected to depend on 
the assumed value for 
$\delta$ in the same manner, and we report all results for 
$\gammin$ with this scaling in mind. For reasonable departures from the assumed values for 
$\Gamma$ and $\delta$, an approximate scaling by $(\Gamma \, \delta)^{-1/2}$ is 
expected to remain even if $\Gamma = \delta$ is not assumed, since the dominant behavior 
of the scattering kernel in all cases includes the photon and electron energies in the 
combination $E_1'/(\gamma^2 \, E_0')$.

\section{RESULTS}
\label{sec:results}

Given that each individual spectrum has only a small number of counts over the 
bandpass of interest and was therefore grouped to have a minimum of only two 
counts per channel, $\chi^2$ is not appropriate as a fitting statistic. Instead, 
the C-statistic implemented in \textsc{xspec} was used. Unfortunately, this 
statistic does not allow for a goodness-of-fit test, in the way the reduced 
$\chi^2$ is commonly used. However, a visual inspection of the fits reveals a 
very good agreement between the data and the model. In any case, the important 
discriminant will be the changes in the fitting statistic with the model 
parameters as well as between models. All subsequent results were obtained 
from simultaneous fits to the 21 individual spectra while fixing the 
model normalization to be the same between the spectra.

\subsection{Phenomenological Fits}
\label{sec:pheno_fits}

We first fit the spectra in the 0.3--7 keV bandpass with a single power law, 
modified by neutral Galactic 
absorption. If $\nh$ is allowed to be free, a best-fit value of 
$(9.1 \pm 0.8) \times 10^{20}$ cm$^{-2}$ is returned, and the best-fitting 
power law energy 
index is $\alpha = 0.76 \pm 0.04$. These values correspond very well with the 
expected values: 
the absorbing column with the value reported by 
\citet{dickey90} 
($9.1 \times 10^{20}$ cm$^{-2}$), and the power law index with both the results 
from earlier investigations of the \axaf{} spectrum of this source 
\citep[][$\alpha = 0.85 \pm 0.08$, ]{chartas00} 
and the radio spectral index 
\citep{schwartz00}.
The same power law index (to within uncertainties) is obtained by fixing \highnh{} 
before fitting. The best-fit C-statistic value is 886.3 for 958 bins.

In the restricted 0.55--7 keV bandpass, the fit with both the power law index and 
the absorbing column free returns $\nh = 3.0 \times 10^{20}$ cm$^{-2}$ (with large 
uncertainties) and $\alpha = 0.63 \pm 0.05$. The C-statistic at the minimum is 
808.2 for 860 bins, but it is clear that the low value for the absorption and 
the resulting fit is spurious, as the restricted bandpass is not very sensitive 
to the absorption column density. 
If \highnh is fixed before fitting, the power law index increases to 
$\alpha = 0.84 \pm 0.02$, once more consistent with previous values. The minimum 
value of the fitting statistic is 817.2 for 860 bins.
In all subsequent fits, the absorbing column is kept fixed at the Galactic value.

The normalization of the power law component (before absorption) is in all cases 
equal to about $(3.1 \pm 0.5) \times 10^{-5}$ photons/cm$^2$/s/keV, corresponding to a flux 
density of ($21 \pm 3$) nJy at
1 keV. This is slightly smaller than, but probably within 1$\sigma$ of, 
previous measurements 
\citep{schwartz00}, 
which could be due to differences in the extraction region or the background subtraction.

\subsection{IC/CMB Model Fits}
\label{sec:iccmb_fits}

\subsubsection{Upper Limit on $\gammin$ from \xray{} Spectrum}

We now wish to investigate whether a better fit can be obtained by letting the 
IC/CMB spectrum cut off before the low-energy end of the bandpass under 
consideration. It is expected that if $\scgammin \lesssim 40$ 
($\delta_{10} = \delta / 10$), the cutoff will 
be too low in energy to affect the 0.3--7 keV bandpass; obviously, this 
limiting value of $\gammin$ will be higher for the 0.55--7 keV bandpass.

We proceed to fit the spectra with the custom IC/CMB models 
developed in \mbox{Section 
\ref{sec:iccmb_model}.}
For the 0.3--7 keV bandpass, the best fit for both the 
\citet{blumenthal70} 
and the 
\citet{aharonian81} 
kernel, and for both the sharp cutoff in the electron distribution or the 
flat segment case (cases (1) and (2) in Figure 
\ref{fig:broadband}), 
is such that the best-fitting $\gammin$ is above the limiting value. In fact, 
plotting the fitting statistic as a function of $\gammin$ leads to Figure 
\ref{fig:gchi_0.3}, which shows that a cutoff due to $\gammin$ appears 
to be detected about 2$\sigma$ statistical confidence for two of the four cases. 
Note that the differences in C-statistic values are equivalent to differences 
in $\chi^2$, such that the same $\Delta\chi^2$ values may be used to determine 
the confidence regions of fitted parameters.

The best-fit values of the fitting statistic and the constraints that we are able 
to place on $\gammin$ are shown in Table 
\ref{tbl:gammin_limits_xray}. 
In all cases, the fit is as good as the unbroken power law fit or better. 
The y-axis in Figure 
\ref{fig:gchi_0.3} 
is normalized to have a $\Delta$(C-statistic) of 0 for the best fit. We expect the 
previous fit to the unbroken power law to be recovered by pushing $\gammin$ below the 
limiting value. In detail, the IC/CMB models to not revert exactly to a power law 
even if $\gammin$ is very low; furthermore, slight differences in the value of the 
fitting statistic are expected near the minimum, given the minimization algorithm 
employed by \textsc{xspec}. The limiting value of the 
fitting statistic as $\gammin \rightarrow 1$ can therefore be slightly different 
from the value obtained for the unbroken power law fit above.

The power law index for the electron distribution is returned in all four cases as 
$s = 2.6 \pm 0.1$, consistent with the measurement of the energy 
index measured for the power law fit ($\alpha = (s - 1) / 2$). 
The index is systematically lower for the case of the flat electron 
distribution, but the difference to the index for the sharp cutoff in the electron 
distribution is not statistically significant. There is almost no degeneracy between 
the power law index and $\gammin$.

As expected, restricting the analysis to the 0.55--7 keV bandpass 
eliminates the minimum in the fitting statistic, and only upper limits 
on $\gammin$ are obtained. The behavior of the fitting statistic as a function 
of $\gammin$ for the restricted bandpass is shown in Figure 
\ref{fig:gchi_0.55}. 
Again, the relevant data on the best fits and on the $\gammin$ constraints are 
included in 
Table 
\ref{tbl:gammin_limits_xray}.

It is worth noting that, contrary to expectations, the upper limits on $\gammin$ from 
the restricted bandpass are as tight as from the full 0.3--7 keV bandpass. The 
reason for this might be that the residuals to the fit in the restricted bandpass 
favor an unbroken power law down to the lowest bins and are very sensitive to 
changes in the shape at the low-energy end, while the residuals in the bins between 
0.3--0.55 keV are more ambivalent about the shape of the model in that range.

\subsubsection{Lower Limit on $\gammin$ from Optical Flux Measurement}

Figure 
\ref{fig:broadband} 
shows the optical and \xray{} measurements of \pks{} as well as the 
predicted IC/CMB spectra based on $\scgammin = 60$ and the best-fit value of 
the power law index $s = 2.6$.
As $\gammin$ becomes smaller, the low-energy extension of the 
model fitted to the \xray{} observations will approach and at some point 
over-predict the observed optical flux 
\citep{schwartz00}. 
This gives us an opportunity to place lower limits on $\gammin$.

The sum of the individual knots' optical flux density is 0.574 $\mu$Jy, and we 
estimate a conservative error on that measurement of 0.1 $\mu$Jy. The 
observed \xray{} flux density at 1 keV is ($21 \pm 3$) nJy. The ratio of 
optical to \xray{} flux densities thus evaluates to $29.0 \pm 9.7$.

The IC/CMB models developed in the previous section were adapted to return 
the ratio of optical to \xray{} flux as a function of $\gammin$. The 
effect of the uncertainty on the power law index measurement was included 
by always calculating the minimum ratio over the confidence region of the 
power law index. 
The calculation of the ratio in the case of the flat electron distribution 
below $\gammin$ requires the caveat that the assumption $\gamma \gg 1$ 
is violated, as $\gamma_0 \sim 3$ for energies corresponding to the 
optical data point.

In all cases, the 
ratio is monotonically increasing with decreasing $\gammin$ as long as 
the departure from power law behavior occurs at an energy below the 
reference \xray{} energy, i.e. when $\scgammin \lesssim 75$ as determined 
by the spectral investigation above. For $\scgammin = 50$, the ratio 
evaluates to less than 10 and is safely below the observed ratios quoted 
above. Lower limits on $\gammin$ are thus obtained by inverting the 
optical/\xray{} ratio vs. $\gammin$ relation and reading off $\gammin$ at 
the appropriate statistical upper limits on the ratio. The lower limits on 
$\gammin$ thus obtained will actually underestimate the true lower limits, 
since the two-sided errors on the optical to \xray{} flux ratio, and not the 
one-sided upper limits, are used.

Table 
\ref{tbl:gammin_limits_optical}
summarizes the lower limits on $\gammin$ thus obtained. Note that at 99\% 
confidence, the lower limit in all cases relaxes to the minimum possible 
value for 
$\gammin$ of 1.

\section{DISCUSSION}
\label{sec:discussion}

Tables 
\ref{tbl:gammin_limits_xray} 
and
\ref{tbl:gammin_limits_optical}
summarize the constraints on $\gammin$ that we are able to place. 
If the full bandpass of 0.3--7 keV is used, marginal evidence for 
a break in the spectrum at around 
$\scgammin = 40$--55 emerges, depending on the model for the 
electron distribution and the kernel for IC scattering. 
Given the uncertainties with the 
low-energy calibration of \axaf , the bins below 0.55 keV might want 
to be excluded from analysis, in which case only upper limits can 
be placed on $\gammin$ from the \xray{} spectrum alone: 
$\scgammin \lesssim 80$.

A spurious break due to $\gammin$ might be detected if the column of 
Galactic absorption used in the fit is an under-estimate. However, 
$\nh$ would need to be increased to about 
$1.4 \times 10^{21}$ cm$^{-2}$ (an increase of over 50\%) to 
make the fitting statistic increase monotonically with 
$\gammin$. The behavior 
of $\nh$ in the vicinity of \pks{} was investigated 
by visually inspecting the IRAS 100 $\mu$m map
\citep{wheelock94}. 
No evidence for any Galactic molecular clouds was found that would average out in 
the large-beam radio surveys resulting in a biased value for the quasar's 
$\nh$ reported in 
\citet{dickey90}. 
The difference between their value and the one reported in 
the Leiden/Argentine/Bonn Survey of Galactic HI 
\citep{kalberla05} 
is only 15\% (the former being the higher value); 
it is therefore unlikely that uncertainties in the 
$\nh$ measurement alone are responsible for the minimum in the fitting statistic. 

\begin{figure}
\includegraphics[angle=270,width=\textwidth]{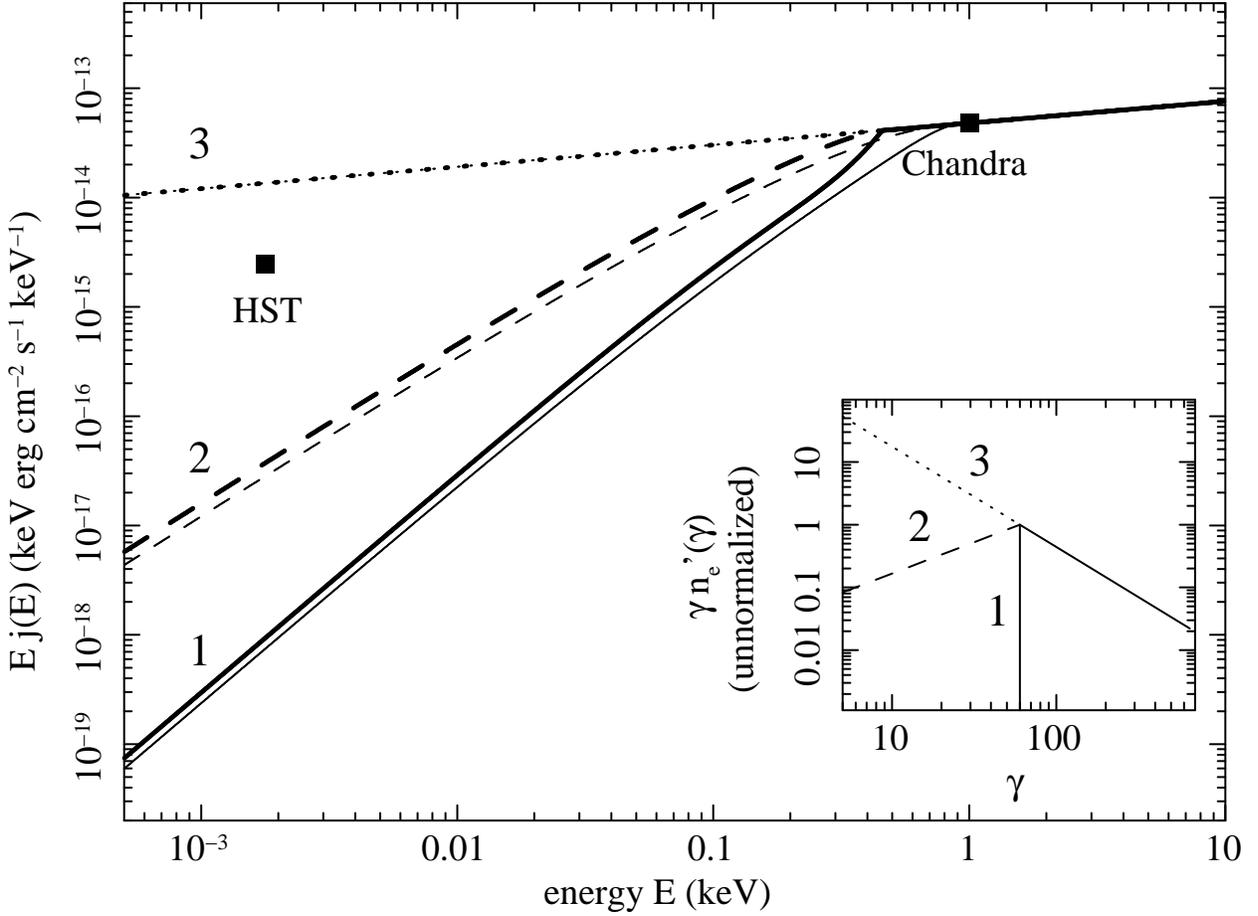}
\caption{
Predicted flux 
of the IC/CMB models developed in \mbox{Section 
\ref{sec:iccmb_model}.} 
The adjustable parameters are set to $\gammin = 60$ and 
$s = 2.6$. 
The inset plot shows the selected electron density distributions, which are all
described as 
a power law with slope $s$ above $\gammin$. Note that the normalization of the 
y axis is arbitrary. The two cases with a modification 
of the spectrum at $\gammin$ are characterized by either 
the step-function cutoff at 
\mbox{$\gammin$ (1)} or the constant density below \mbox{$\gammin$ (2).} \mbox{Case 3} results when 
$\gammin$ is chosen sufficiently low such that the cutoff in the photon spectrum 
moves outside its bandpass. 
The spectra based on the 
\citet{aharonian81} 
kernel are shown as bold lines, the ones based on the 
\citet{blumenthal70}
kernel as thin lines.
The spectra for the two models coincide for case (3). 
Both sets of spectra have been normalized to the observed \xray{} flux. The optical 
data point is also plotted and is violated by the models without a cutoff in the 
electron distribution. The allowable range of $\gammin$, such that the cutoff occurs 
below the \xray{} data point, and the optical flux from the IC/CMB model does not over-predict 
the optical detection, is approximately 5--80.
\label{fig:broadband}}
\end{figure}

In Figure 
\ref{fig:broadband}, 
it can be seen that, for the same value of $\gammin$, the 
point at which the spectrum deviates from power law shape is 
at a lower energy for the 
\citet{aharonian81} 
kernel for inverse Compton scattering compared to the 
\citet{blumenthal70} 
kernel, by about a factor of 3.
This results in tighter lower limits 
on $\gammin$ for the 
\citet{aharonian81} 
kernel, but correspondingly less-strict upper limits. 

The 
\citet{aharonian81} 
kernel is the more appropriate one for the situation of a 
jet moving relativistically through a homogeneous and isotropic 
distribution of seed photons, as is the case for the cosmic 
microwave background. The comparison to the other kernel can 
serve to illustrate the order of magnitude of the uncertainty 
introduced by the choice of the inverse Compton scattering formalism.

We only investigate two possible shapes of the 
cutoff at $\gammin$: a sharp cutoff such that there are no electrons 
below $\gammin$ (case 1 in Figure 
\ref{fig:broadband}), 
and an electron distribution that has a constant density below 
$\gammin$ (case 2 in the same figure).
The real shape of the cutoff is likely to be more complicated than 
in either of these models, but these two distributions bracket the 
range of expected real distributions, where the cutoff is probably 
less sharp than in case 1, but the electron distribution does cut 
off more quickly than to a constant distribution below $\gammin$. 
As expected, if the cutoff is sharp, the fitting statistic increases 
rapidly as $\gammin$ increases, while the increase is slower for the 
case of the flat distribution below $\gammin$, since the \xray{} 
spectrum in this case exhibits a correspondingly milder cutoff.

With $\scgammin \lesssim 80$, and assuming $\Gamma = \delta$, the 
equipartition magnetic field strength 
for a jet composed of hot electrons and cold protons is $\approx 10 \mu$G 
\citep{dermer04}, 
and the radio spectrum is expected to extend unbroken down into the 
hundreds of kHz 
\citep{schwartz07}.

\citet{georganopoulos05}, 
based on the work of 
\citet{dermer04}, 
calculate the kinetic power in the \pks{} jet as a function of $\gammin$ for 
the two extreme 
cases of a hadronic jet (equal numbers of electrons and [cold] protons) 
and a purely leptonic jet, assuming 
that IC/CMB is the dominant emission mechanism for the \xray s, and 
that $\Gamma = \delta$. Our limit of 
$\scgammin \lesssim 80$ is consistent with the ranges in $\gammin$ 
considered in their paper, for both the leptonic and hadronic jet, and 
allows for a jet power of around $5 \times 10^{46}$ erg s$^{-1}$ for 
a hadronic jet and $< 10^{46}$ erg s$^{-1}$ for a leptonic jet. 
Estimations of the jet power in earlier work, such as 
\citet{tavecchio00} 
($L = 3 \times 10^{48}$ erg s$^{-1}$ for $\gammin = 10$), 
\citet{celotti01} 
($L = 8 \times 10^{47}$ erg s$^{-1}$ for $\gammin = 10$--20), or 
\citet{uchiyama05} 
($L = 9 \times 10^{46}$ erg s$^{-1}$ for $\gammin = 20$) 
are likely to be too high, given that $\gammin$ can in fact be higher 
than the assumed values without violating the \xray{} spectral data. 
Note also that the limits $10 \lesssim \gammin \lesssim 40$ in 
\citet{uchiyama05} 
are too restrictive in light of our results.

For the adopted jet angle with respect to the line of sight of 
$6.4{^\circ}$, the deprojected length of the \xray{} 
jet is on the order of 900 kpc. With the estimated magnetic field 
strength of 10 $\mu$G, the cooling time (taking both IC/CMB and 
synchrotron losses into account) for electrons with 
$\gamma \lesssim 8,500$ is larger than the travel time along 
the jet
\citep[Equation D1 in][]{stawarz04}.
It would be possible therefore to produce these electrons 
in the central core and propagate them at the 
bulk velocity to the sites of \xray{} emission.

The high contrast between the \xray{} knot and the inner jet in 
\pks{} argues 
against the presence of a significant population of electrons with 
Lorentz factors in the range from $\approx$100--1000 in the inner jet 
region at the current epoch. Whether the morphology 
is due to modulated jet activity or to localized 
particle acceleration sites remains unresolved.

\section{CONCLUSION}
\label{sec:conclusion}

As far as we are aware, this work represents the first attempt 
at placing confidence limits on the low-energy electron 
distribution cutoff in a jet based on a fit to its \xray{} 
spectrum. Usually, papers investigating the broad-band spectra 
of jet sources simply quote a value for $\gammin$ that happens 
to make the model pass through the measured data points.
It is very hard to conduct observations in the radio that are 
able to constrain $\gammin$ 
\citep{gopalkrishna04},
especially for the kpc-scale jets, where the frequency at which 
the cutoff due to $\gammin$ would manifest itself is below the 
synchrotron self-absorption frequency. 
Thus, the 
soft \xray{} spectrum as well as any measured 
optical fluxes are the best tools to shed light on the 
behavior of the low-energy end of the electron distribution.

We find that $\scgammin \lesssim 80$, which is significantly higher 
than the value of 10--20 that has previously been assumed in broad-band 
spectral modeling of 
this source. The kinetic power requirement is therefore lessened, 
but the questions of jet composition and \xray{} emission mechanism 
are not addressed conclusively with this finding.

The present work did not consider other proposed emission 
mechanisms for the \xray{} spectrum, such as 
direct synchrotron or synchrotron self-Compton. 
Based on the above considerations, however, IC/CMB remains a viable 
model for the \xray{} emission in \pks .

\acknowledgements
This research was supported in part by NASA Contract 
\mbox{NAS8-39073} to the 
\axaf{} \xray{} Center and CXC grant 
\mbox{GO3-4120X} to SAO, as well as by 
the Department of Energy Contract 
\mbox{DE-AC02-76SF00515} 
to the Stanford Linear Accelerator Center. 
The data reduction made use of the \axaf{} Interactive 
Analysis of Observations tools 
\mbox{(http://cxc.harvard.edu/ciao),} 
\mbox{version 3.3,} 
\mbox{CALDB version 3.2.4.} 
Spectral fits were obtained in \textsc{xspec} 
\citep{arnaud96}.
We wish to thank the anonymous referee for many helpful 
comments, and Mark Bautz and Catherine Grant for assistance 
understanding the low-energy response of \axaf . Further 
thanks to Greg Madejski and \L ukasz Stawarz for 
fruitful discussions at all stages of this work.

\clearpage

\input{tab1.tex}
\input{tab2.tex}
\input{tab3.tex}

\clearpage

\begin{figure}
\includegraphics[angle=270,width=\textwidth]{f2.eps}
\caption{Confidence limits on $\gammin$ for the 
IC/CMB model fits, obtained from the 0.3--7 keV \xray{} spectrum. 
The bold lines are for the 
\citet{aharonian81} 
kernel, the thin lines for the 
\citet{blumenthal70} 
kernel.
In both cases, the solid line is for the electron distribution with 
the sharp cutoff at $\gammin$ (case (1) in Figure 
\ref{fig:broadband}), and the dashed line for the constant segment 
below $\gammin$ (case (2) in Figure 
\ref{fig:broadband}).
The three dotted lines mark the $\Delta$(C-statistic) = 1.00, 2.71 and 
6.63 levels corresponding to the 68.3, 90 and 99\% two-sided 
confidence limits on $\gammin$.
\label{fig:gchi_0.3}}
\end{figure}

\begin{figure}
\includegraphics[angle=270,width=\textwidth]{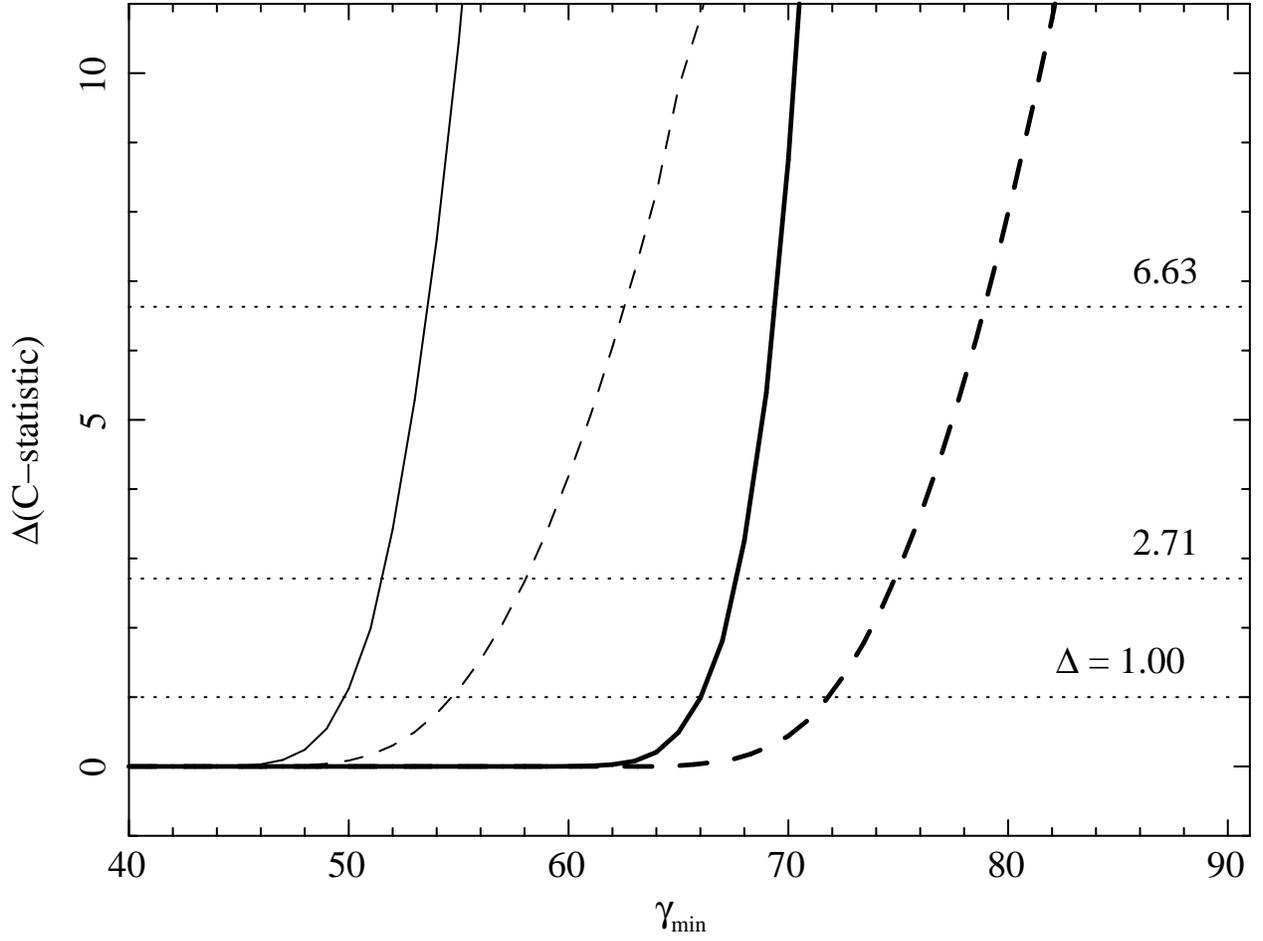}
\caption{Confidence limits on $\gammin$ for the 
IC/CMB model fits, obtained from the \xray{} spectrum using the 
restricted 0.55--7 keV bandpass. The same conventions as for Figure 
\ref{fig:gchi_0.3} 
apply.
\label{fig:gchi_0.55}}
\end{figure}

\end{document}

%% file: tab1.tex
\begin{deluxetable}{rr|rr}
\tablewidth{0pt}
\tablecaption{\axaf{} \textsc{acis-s} Observations of \pks{} Used 
in this Investigation\tablenotemark{a}
\label{tbl:observations}}
\tablehead{
\colhead{ObsID} & \colhead{Exposure} & 
\colhead{ObsID} & \colhead{Exposure}\\
 & \colhead{Time (ks)} &
 & \colhead{Time (ks)}}

\startdata

 472 & 5.52 &  1062 &  0.62\\
 473 & 3.54 &  1063 &  1.22\\
 474 & 4.64 & 62549 &  6.25\\
 475 & 4.64 & 62550 &  5.20\\
 476 & 2.39 & 62551 &  5.32\\
1051 & 1.02 & 62552 &  5.07\\
1052 & 1.02 & 62553 &  4.88\\
1055 & 2.01 & 62554 & 11.22\\
1056 & 1.73 & 62555 &  4.91\\
1058 & 1.51 & 62556 &  4.84\\
1060 & 0.62 & \textbf{Total} & \textbf{78.20}\\

\enddata

\tablenotetext{a}{More information on the individual observations 
is available in Table 1 of 
\citet{chartas00}.}

\end{deluxetable}

%% file: tab2.tex
\begin{deluxetable}{llcccc}
\tablewidth{0pt}
\tabletypesize{\footnotesize}
\tablecaption{Constraints on $\gammin$ Obtained from the \xray{} Spectrum
\label{tbl:gammin_limits_xray}}

\tablehead{
  \colhead{\xray{}} &
  \colhead{IC/CMB kernel\tablenotemark{a}} & 
  \colhead{e$^-$} & 
  \colhead{best-fit} &
  \colhead{$\scgammin$} & 
  \colhead{$\scgammin$}\\
  \colhead{bandpass}&
  \colhead{or power law model} &
  distr.\tablenotemark{b} &
  C-statistic &
  (1$\sigma$) &
  (99\%)\\
}

\startdata
0.3--7 keV  & power law            & \nodata  & 886.3 &            &            \\
            & BluGou               & case (1) & 879.4 & 49 $\pm$ 2 & 49 $\pm$ 6 \\
            &                      & case (2) & 881.5 & 57 $\pm$ 4 & $<67$      \\
            & AhaAto               & case (1) & 881.8 & 63 $\pm$ 2 & $<67$      \\
            &                      & case (2) & 880.2 & 72 $\pm$ 3 & 72 $\pm$ 10\\
\\
0.55--7 keV & power law            & \nodata  & 817.2 &            &            \\
            & BluGou               & case (1) & 818.6 & $<50$      & $<54$      \\
            &                      & case (2) & 817.2 & $<55$      & $<63$      \\
            & AhaAto               & case (1) & 817.3 & $<67$      & $<70$      \\
            &                      & case (2) & 817.3 & $<73$      & $<79$      \\
\enddata

\tablenotetext{a}{BluGou = \citet{blumenthal70}, AhaAto = \citet{aharonian81}}
\tablenotetext{b}{Cases (1) and (2) refer to the different electron 
distributions in Figure 
\ref{fig:broadband}.}

\end{deluxetable}

%% file: tab3.tex
\begin{deluxetable}{lccc}
\tablewidth{0pt}
\tablecaption{Lower Limits on $\gammin$ Obtained from Optical and \xray{}
Flux Measurement
\label{tbl:gammin_limits_optical}}

\tablehead{
  \colhead{IC/CMB kernel\tablenotemark{a}} & 
  \colhead{e$^-$} & 
  \colhead{$\scgammin$} & 
  \colhead{$\scgammin$}\\
  &
  distr.\tablenotemark{b} &
  (1$\sigma$) &
  (99\%)\\
}

\startdata
BluGou               & case (1) & $>4.2$ & $>1$\\
                     & case (2) & $>6.1$ & $>1$\\
AhaAto               & case (1) & $>4.8$ & $>1$\\
                     & case (2) & $>7.7$ & $>1$\\
\enddata

\tablenotetext{a}{BluGou = \citet{blumenthal70}, AhaAto = \citet{aharonian81}}
\tablenotetext{b}{Cases (1) and (2) refer to the different electron 
distributions in Figure 
\ref{fig:broadband}.}

\end{deluxetable}

%% file: ms.bbl
\begin{thebibliography}{}

\bibitem[Aharonian \& Atoyan(1981)]{aharonian81}
Aharonian, F. A. \& Atoyan, A. M. 1981, Ap\&SS, 79, 321

\bibitem[Arnaud(1996)]{arnaud96}
Arnaud, K. A. 1996, in ASP Conf. Ser. 101, 
Astronomical Data Analysis Software and Systems V, 
ed. G. Jacoby \& J. Barnes (Provo, UT: Astronomical 
Society of the Pacific), 17

\bibitem[Blumenthal \& Gould(1970)]{blumenthal70}
Blumenthal, G. R. \& Gould, R. J. 1970, RevModPhys, 42, 237

\bibitem[Celotti, Ghisellini \& Chiaberge(2001)]{celotti01}
Celotti, A., Ghisellini, G., \& Chiaberge, M. 2001, MNRAS, 321, L1

\bibitem[Chartas \etal (2000)]{chartas00}
Chartas, G. \etal{} 2000, ApJ, 542, 655

\bibitem[Dermer \& Atoyan(2004)]{dermer04}
Dermer, C. D. \& Atoyan, A. 2004, ApJ, 611, L9

\bibitem[Dickey \& Lockman(1990)]{dickey90}
Dickey, J. M. \& Lockman, F. J. 1990, ARAA, 28, 215

\bibitem[Felten \& Morrison(1966)]{felten66}
Felten, J. E., Morrison, P. 1966, ApJ, 146, 686

\bibitem[Fixsen \etal (1996)]{fixsen96}
Fixsen, D. J., Cheng, E. S., Gales, J. M., Mather, J. C., Shafer, R. A., 
\& Wright, E. L. 1996, ApJ, 473, 576

\bibitem[Georganopoulos \etal (2005)]{georganopoulos05}
Georganopoulos, M., Kazanas, D., Perlman, E., \& Stecker, F. W. 2005, ApJ, 625, 656

\bibitem[Gopal-Krishna, Biermann \& Wiita(2004)]{gopalkrishna04}
Gopal-Krishna, Biermann, P. L., \& Wiita, P. J. 2004, ApJ, 603, L9

\bibitem[Harris \& Krawczynski(2002)]{harris02}
Harris, D. E. \& Krawczynski, H. 2002, ApJ, 565, 244

\bibitem[Kalberla \etal (2005)]{kalberla05}
Kalberla, P. M. W., Burton, W. B., Hartmann, Dap, Arnal, E. M., Bajaja, E., 
Morras, R., \& Pöppel, W. G. L. 2005, A\&A, 440, 775

\bibitem[Lovell \etal (2000)]{lovell00}
Lovell, J. E. J. \etal{} 2000, in Astrophysical Phenomena Revealed by Space 
VLBI, ed. H. Hirabayashi, P. G. Edwards, \& D. W. Murphy (Sagamihara, Japan: 
ISAS), 215

\bibitem[Marshall \etal (2005)]{marshall05}
Marshall, H. L. \etal{} 2005, ApJS, 156, 13

\bibitem[Press \etal (1992)]{Press}
Press, W. H., Teukolsky, S. A., Vetterling, W. T., Flannery, B. P. 1992, 
Numerical Recipes in C (Cambridge, U.K.: Cambridge Univ. Press)

\bibitem[Rybicki \& Lightman(1979)]{rybicki}
Rybicki, G. B. \& Lightman, A. P. 1979, Radiative Processes in Astrophysics 
(New York, NY: John Wiley \& Sons)

\bibitem[Sambruna \etal (2002)]{sambruna02}
Sambruna, R. M. \etal{} 2002, ApJ, 571, 206

\bibitem[Sambruna \etal (2004)]{sambruna04}
Sambruna, R. M. \etal{} 2004, ApJ, 608, 698

\bibitem[Sambruna \etal (2006)]{sambruna06}
Sambruna, R. M. \etal{} 2006, ApJ, 641, 717

\bibitem[Savage, Browne, \& Bolton(1976)]{savage76}
Savage, A., Browne, I.W.A., \& Bolton, J. G. 1976, MNRAS, 177, 77P

\bibitem[Schwartz \etal (2000)]{schwartz00}
Schwartz, D. A. \etal{} 2000, ApJ 540, L69

\bibitem[Schwartz \etal (2006)]{schwartz06}
Schwartz, D. A. \etal{} 2006, ApJ 650, 592

\bibitem[Schwartz(2007)]{schwartz07}
Schwartz, D. A. 2007, in Revista Mexicana de Astronom\'\i a y Astrof\'\i sica 27, 
Triggering Relativistic Jet, ed. W. H. Lee \& E. Ram\'\i rez-Ruiz 
(Mexico City, Mexico: UNAM), 102

\bibitem[Siemiginowska \etal (2002)]{siemiginowska02}
Siemiginowska, A., Bechtold, J., Aldcroft, T. L., Elvis, M., Harris, D. E., 
\& Dobrzycki, A. 2002, ApJ, 570, 543

\bibitem[Siemiginowska \etal (2003)]{siemiginowska03}
Siemiginowska, A., Smith, R. K., Aldcroft, T. L., Schwartz, D. A., Paerels, F., 
\& Petric, A. O. 2003, ApJ, 598, L15

\bibitem[Stawarz \etal (2004)]{stawarz04}
Stawarz, \L., Sikora, M., Ostrowski, M., Begelman, M. C. 2004, ApJ, 608, 95

\bibitem[Stawarz \etal (2005)]{stawarz05}
Stawarz, \L., Siemiginowska, A., Ostrowski, M., \& Sikora, M. 2005, ApJ, 626, 120

\bibitem[Tavecchio \etal (2000)]{tavecchio00}
Tavecchio, F., Maraschi, L., Sambruna, R. M., Urry, C. M. 2000, ApJ, 544, L23

\bibitem[Tingay \etal (1998)]{tingay98}
Tingay, S. J. \etal{} 1998, ApJ, 497, 594

\bibitem[Uchiyama \etal (2005)]{uchiyama05}
Uchiyama, Y. \etal{} 2005, ApJ, 631, L113

\bibitem[Wheelock \etal (1994)]{wheelock94}
Wheelock, S.-L. \etal{} 1994, IRAS Sky Survey Atlas, JPL Publication 94-11 
(Pasadena: Jet Propulsion Laboratory)

\end{thebibliography}
